# Electric Field-induced Charge Transport in Redox-active Molecular Junctions

Ritu Gupta[1], Shapath Bhandari[1], Savas Kaya[2], Konstantin P. Katin[3] and Prakash Chandra Mondal[1,*]

[1]Department of Chemistry, Indian Institute of Technology Kanpur, Uttar Pradesh-208016, India

[2]Department of Pharmacy, Faculty of Science, Cumhuriyet University, Sivas 58140, Turkey

[3]Institute of Nanotechnologies in Electronics, Spintronics and Photonics, National Research Nuclear University "MEPhI", Moscow 115409, Russia

**Abstract**:

The formation of well-defined three-dimensional (3D) redox-active molecular nanostructures at the electrode surfaces may open additional routes to achieve higher conductance in molecular junctions (MJs). We report here experimental and theoretical charge transport analysis on electroactive ruthenium(II)-tri(phenanthroline) [Ru(Phen)$_3$]-based molecular junctions covalently grown on patterned ITO electrode. Thicknesses of the molecular layers are varied between 4 to 13 nm, thanks to the potential-driven electrochemical technique to achieve it. A thin layer of Al was deposited on top contact over ITO/ Ru(Phen)$_3$ to fabricate large-area solid-state molecular junctions with a stacking configuration of ITO/[Ru(Phen)$_3$]$_{4nm, 10nm, 13nm}$/Al. The electrified molecular junctions show LUMO-mediated electron-driven resonant charge conduction with attenuation in conductance as a function of the length of Ru(Phen)$_3$ layers ($\beta$ = 0.48 to 0.60 nm$^{-1}$). Molecular junctions consisting of 4 nm Ru(Phen)$_3$ layers follow quantum tunneling, while the thicker junctions (10, and 13 nm) follow Poole-Frenkel and electric-field induced charge conduction. Considering the energy level of frontier molecular orbitals, Fermi energy of ITO, and Al contact, a mechanism of symmetric current-voltage features with respect to the bias-polarity is predicted. The present work describes a simple, controllable, low-cost, and versatile approach to fabricating 3D molecular assembly for mimicking conventional electronic functions.

**Introduction**

The prime core of molecular electronics is to cognize and regulate charge transport mechanisms at nanoscale molecular junctions with varied molecular structures, compositions, and electrode combinations.[1–4] Toward this goal, a forward-looking approach is to assemble either a single molecule or a few molecules between two electrical conductors, which is popularly known as molecular electronics.[5–7] Charge conduction with respect to an external bias that is applied to molecular junctions such as metal/molecules/metal can get influenced by factors.[8–11] The factors include hybridization, electronic coupling to the electrode/molecule(s) interfaces, electrode composition, and Fermi energy, the thickness of the molecular layers, packing and orientation on the electrode surface, and frontier orbital energies governing charge transport phenomena.[12–15] Significant progress has been made with molecular junctions consisting of organic molecules as testbeds to reveal underline charge transport characteristics.[16–18] Besides, many molecular junctions that are made of metal centers containing π-conjugated backbone are able to exhibit long-range charge transport due to either less band gap between HOMO and LUMO, and involvement of redox centers, and subsequently, their current-voltage response is almost independent of length, thus yielding low attenuation factors.[19–22] Ruthenium-polypyridyl complexes are considered model systems for investigating photophysical studies, and find wider applications in redox, catalytic, sensing, and biological properties.[23–27] However, not much work is reported with ruthenium-polypyridyls in molecular junctions for understanding charge transport across the films with varied structures, configurations, and compositions.[28–31] Besides, all the reports, among the chemical approaches focus on thiolated self-assembled monolayers (SAMs), silane-based coupling layers, phosphonic linkers, or single and bis aryl diazonium salts for electrochemically (E-Chem) grafting the molecular layer on the various bottom electrodes ranging from Au, doped Si, Cu, Co, Ni, ITO, carbon contacts, hence providing only one vertical channel for charge transport in the molecular junctions.[20,32,33] Electrochemical reduction method in creating molecular layers of diverse structures, composition, and desired thickness is much advantages and this technique is gaining gigantic attention to form robust bond at the (working) electrode-molecules interfaces.[34–38] This technique facilitates growing the molecular films much faster over the self-assembly methods as the process happens via forming reactive radials near the working electrode surface via electroreduction of corresponding aryl diazonium salts.[39–41] Thus, creating varied thickness of the molecular layers keeping the structure nearly identical is the crucial condition for comparing thickness-dependent charge transport phenomena. Three-dimensional (3D) coordination-based molecular assemblies offer fascinating features including optical, redox, electrochromic, logic-gate, linear vs. exponential films growth are mostly grown via layer-by layer (LbL) method using stepwise coordination between metal ions, and organic linkers.[27,42–44] Incorporating such 3D molecular assemblies where more no. of redox centres are accumulated, in a molecular junction for understanding electric-field induced charge transport at systematic layer thickness variation would help to mimic conventional electronic functions. The present work deals with $Ru(Phen)_3(PF_6)_2$ complex as an active circuit element grown in a three-dimensional architectures on patterned indium tin oxide (ITO) followed by aluminium (Al) top contact deposition to complete electronic circuit. Three-dimensional network of Ru-complexes provides lateral hopping and increases the number of redox-active Ru(II) centers, overall assisting in enhancing the conductance of the device.

**Result & discussion**

We choose here popular ligand framework which is 1,10-phenanthroline (or simply phen) very similar to bidentate chelating ligand 2,2'-bipyridine (bipy) for coordinating it with Ru(II). We have synthesized Ru(II) complex bearing three 5-amino-1,10-phenanthroline (Phen-NH$_2$) ligands by following previously reported literature.[45] Detailed synthesis procedure and characterization of Ru(II) complex are discussed in supporting information (SI) (**Figure S1-S9**). The Ru(Phen)$_3$(PF$_6$)$_2$ was electrochemically grafted on custom-made patterned ITO substrate by using in situ diazonium generation. Reduction of tris-diazonium generated radicals at three different sites; hence facilitated the 3-dimensional growth of Ru(Phen)$_3$ oligomers on ITO working electrode (**Figure 1a**). A broad reduction peak at − 0.15 V vs. Ag/AgNO$_3$ denotes the reduction of diazonium to radicals (**Figure 1b**). The reduction in current density on repeated cycles suggests that a partially insulating film is grafted on the ITO surface.[46] Details of E-Chem grafting parameters for creating molecular layers of different thicknesses are discussed in supporting information (**Figure S10-S12, Table S1**). A clear well-ordered triangular grain structure across the surface and no dark patches confirms the homogeneity of the film (**Figure 1c-d**). The root-mean-square (RMS) roughness of the Ru(Phen)$_3$ film was also calculated from AFM measurement and was found to be ~ 2.8 nm, close to bare ITO. Next, the thickness of grafted layers was also determined from AFM 'scratching' technique (**Figure S13-S14**). By varying electrochemical parameters such as no. of CV scans, increased electrochemical potential window, the thickness was successfully varied from ~ 4 nm to ~ 13 nm. The UV-vis absorbance spectra of the thin films show two peaks at 365 nm and 570 nm, corresponding to π-π$^*$ transition, metal-to ligand charge transfer band (**Figure S15**). Also, broadening in the absorption spectra confirms the grafting of the metal complex on ITO.[47]

Cyclic voltammogram of the Ru(Phen)$_3$ films show a characteristic one electron oxidation peak due to Ru(II) to Ru(III) forming at 1.36 V vs. Ag/AgNO$_3$ and a reduction peak due to Ru(III) to Ru(II) at 1.08 V vs. Ag/AgNO$_3$ (**Figure S16a**).[47] While the reduction peaks around -1.04 V and -1.40 V vs. Ag/AgNO$_3$ arose due to the reduction of phenanthroline ligands (**Figure S16b**). X-ray photoelectron spectroscopy (XPS) survey spectra denoted all the expected elements peaks, including C1s, Ru3d, O1s, N1s, F1s, and P2p for Ru(Phen)$_3$ grafted ITO (**Figure S17,18a**). x-ray photoelectron spectrum shows a characteristic peak at 280 eV assigned to the Ru(II) 3d$_{5/2}$, and the corresponding Ru(II) 3d$_{3/2}$ is masked under the C1s peak (**Figure S18b**).[28] A new peak centred at 531.5 eV in the O1s spectra obtained corresponding to In-O-Ph, suggesting the binding of formed radicals to the hydroxylated ITO surface. Details of XPS data are discussed in supporting information (**Figure S17-S18, Table S2**).

After ensuring the formation of electrochemically generated 3D-molecular assemblies, ta top Al electrode was deposited for fabricating molecular junctions, as shown schematically in **Figure 2a, S19**. Current density vs. bias voltage (j-V) plots for Ru(Phen)$_3$ MJs with three different thicknesses (~ 4 nm, ~ 10 nm, and ~ 13 nm) are provided in **Figure 2b-d**. The j-V curves shown here are the average of individual MJs shown in bold colour with the typical error bar in each case. The j-V responses for all three thicknesses are non-linear and slight asymmetric with respect to bias polarity. Despite using electrodes with different work functions, the absence of rectification is quite surprising here. The difference in current density (j) by the orders of magnitude for three different thicknesses of Ru(Phen)$_3$ oligomers clearly depicts a strong dependence on molecular lengths, as shown in semilog plots (**Figure 2e**). For instance, on changing the thickness from ~ 4 to ~ 13 nm, the measured j value at + 1 V decreases nearly 80 times. This "thickness signature" is usually illustrated in MJs. For Ru(Phen)$_3$ 3D molecular assembly, the linear dependence of Ln j with molecular thickness from ~ 4 to ~ 13 nm at a particular bias indicates that the simplified Simmons equation (i) applies over such a large thickness range, which suggests tunneling current decreases exponentially with the molecular thickness. Here, j is current density at particular bias and β is tunneling attenuation factor.[1]

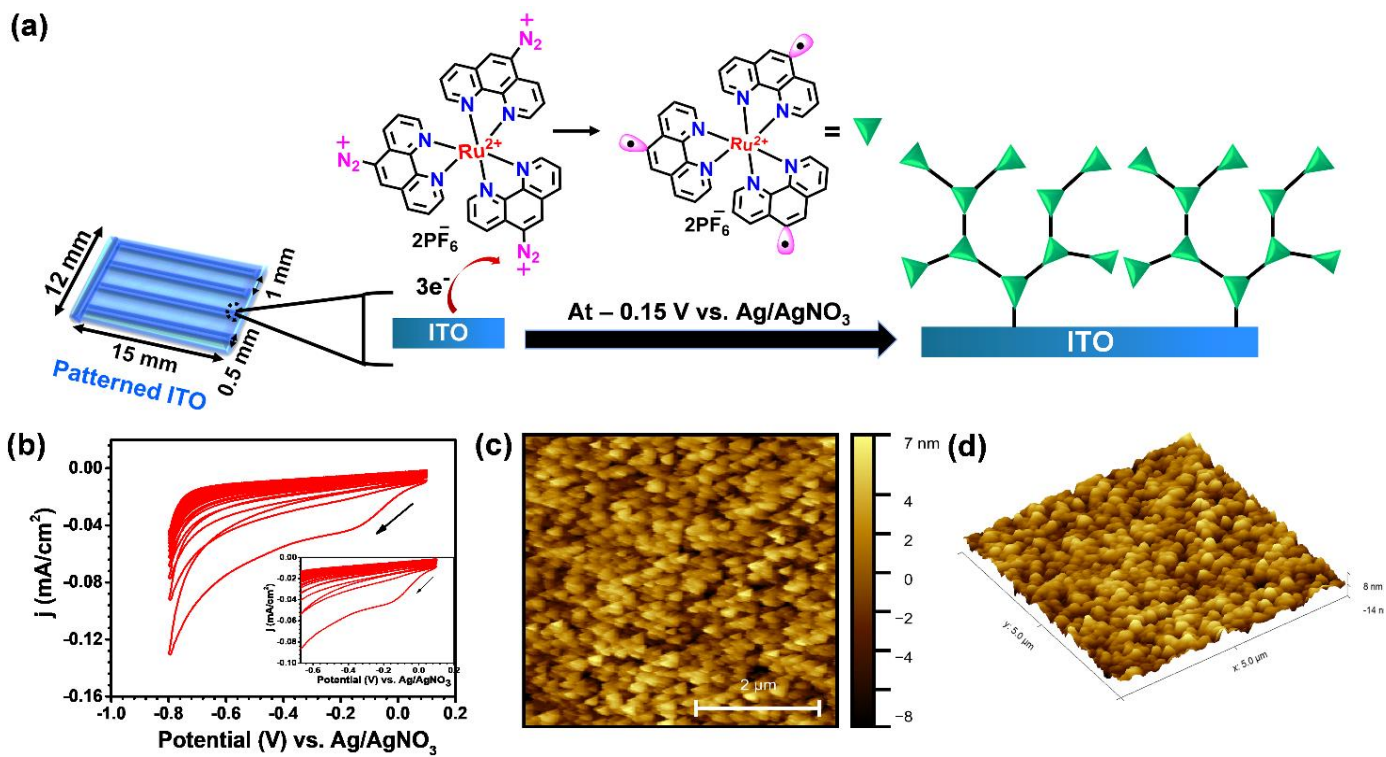

Figure 1. (a) Proposed schematic of E-Chem grafting of in-situ generated diazonium cations of [Ru(Phen-NH$_2$)$_3$] (PF$_6$)$_2$ complex on patterned ITO substrate. (b) CV of E-Chem reduction of 0.5 mM in-situ generated diazonium cations of [Ru(Phen-NH$_2$)$_3$] (PF$_6$)$_2$ complex at 100 mV/s scan rate up to 15 scans. (c,d) 2d and 3d non-contact mode AFM image of Ru(Phen)$_3$ thin film on ITO.

$$j = j_0 e^{-\beta d} \qquad (i)$$

**Figure 2f** shows the Ln j vs. d (β plot) for Ru(Phen)$_3$ at several bias values to demonstrate the dependence of current density on molecular length. β values less than 1 nm$^{-1}$ are generally observed for redox-active MJs. At V = + 0.25 V and + 0.5 V, the β value was found to be 0.60 nm$^{-1}$ and 0.54 nm$^{-1}$, which decreases to 0.48 nm$^{-1}$ at 1 V bias, thus fitting in the desired range of redox-active MJs. A similar β value is also reported for molecular

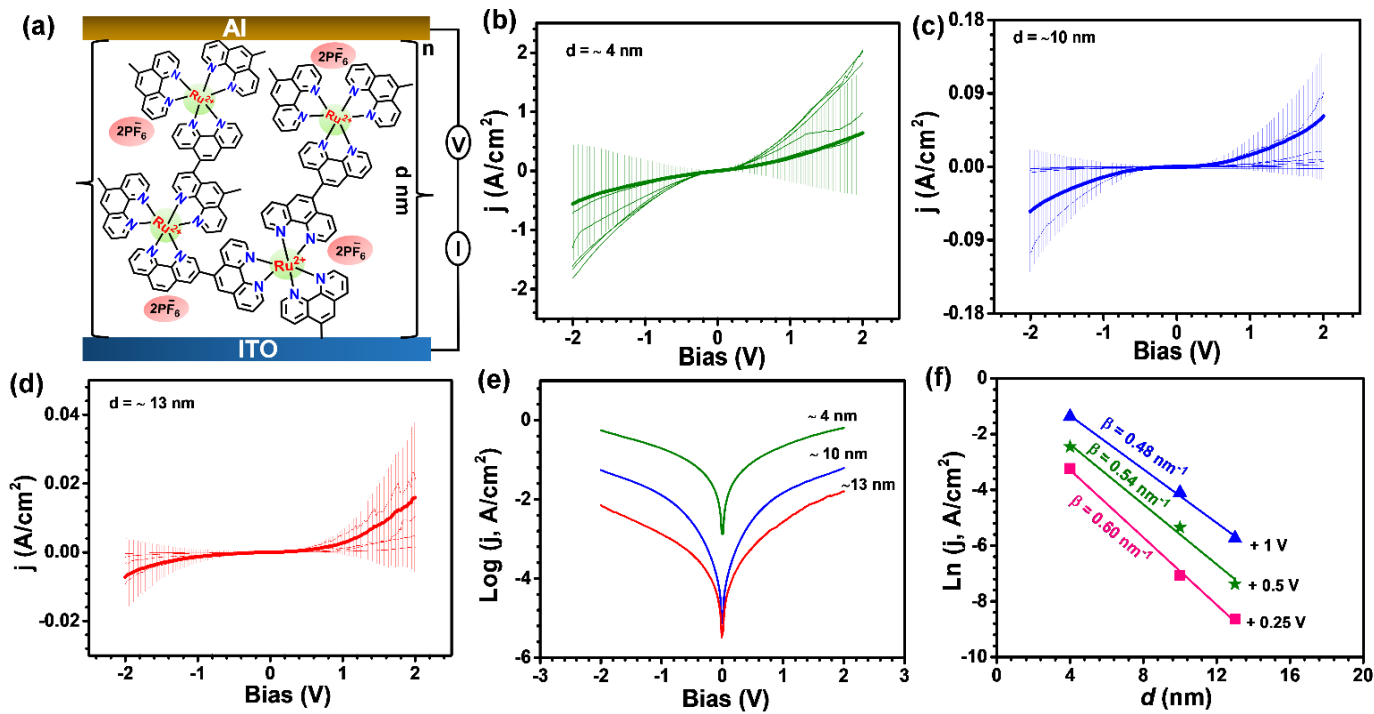

Figure 2. (a) Schematic illustration of the large area Ru(Phen)$_3$ MJs. (b) jV curves for Ru(Phen)$_3$ MJs with the molecular thickness of 4 nm, (c) 10 nm, and (d) 13 nm. (e) Average semilog plot from the individual MJs. (f) Ln j vs. d plot at different bias voltage.

Junctions of Ru(II)-polypyridyl, but at a relatively higher bias = 3 V.[28] The result shown here clearly demonstrates that the β value in Ru(Phen)$_3$ MJs is very low (0.48 nm$^{-1}$), and there is no variation in the slope of the β plot suggesting no transition in the two transport regimes. Hence, we can exclude the coherent off-resonant tunnelling as the dominant transport mechanism in Ru(Phen)$_3$ MJs with thicknesses ~ 4 to ~ 13 nm, which mainly applies to below 5 nm thickness of molecular layers. Further, plots of the first derivative of conductance (dj/dV) vs. bias (V) was made to illustrate the local density of states (LDOS) for all three different thickness of the MJs (**Figure 3a, S20-21**). The dj/dV curve was plotted from the average j-V curve of the MJs for all cases. The peaks suggest that the extra conduction states in the system. In agreement with j-V curves, dj/dV curves illustrate the slight asymmetric profile for asymmetric MJs. In addition to tunneling and hopping, there are several other classical mechanisms involved in MJs, including field ionization (i.e., Fowler–Nordheim (FN) tunneling), Poole–Frenkel (PF) transport between Coulombic traps, and space charge limited conduction (SCLC). Few of the charge transport mechanisms are electric field dependent rather than the applied bias voltage. Previously reported bisthienylbenzene (BTB) and Ru(bpy)$_3$ MJs show E-field-dependent charge conduction.[29,48] For Ru(Phen)$_3$ MJs, a non-linear plot of ln (J/E$^2$) vs. 1/E (i.e., an FN plot), where E is the electric field, concludes that the FN mechanism is not consistent with applied bias ranges (Figure S22). **Figure 3b** shows the Ln j vs. E plot from the similar average j-V curves for d = 4 – 13 nm. As d increases from 4 nm, the Ln j vs. E curve nearly superimposes, suggesting that for higher thickness, charge transport is E-field dependent rather than applied bias.[48] The PF charge transport mechanism expected to show a linear plot of Ln (j/E) vs. E$^{1/2}$; similar curves are shown in Figure 3c, S23.[49] The linear behaviour in the plot for d

> 4 nm indicates that the PF conduction mechanism may be valid for d > 4 nm Ru(Phen)$_3$ layers. SCLC is another conduction mechanism that occurs when the density of injected charge carriers increases the density of free thermal carriers already present in the device. Considering PF electric field dependence and molecular layers with traps, the SCLC can be approximated as equation (ii), where $\mu_0$ is zero field mobility, E is electric field, and $\gamma$ is field enhancement factor.[50]

$$j = \frac{9}{8} \epsilon\epsilon_0 \mu_0 \frac{V^2}{d^3} e^{(0.891\gamma\sqrt{E})} \qquad (ii)$$

In principle, in the presence of traps, the SCLC current follows the power law form; $j_{SCLC} = V^n$. Hence, Log j vs. Log V plot was plotted from average j-V plots for all thickness, which shows three regions with varying slopes (n) (**Figure 3d, S24**). At low bias voltage in region I, the number of injected charge carriers did not exceed the free thermal charge carriers, thus following the Ohms law $j = \frac{qn\mu V}{d}$ with a slope n = ~ 1 for all thickness ranges. At higher bias voltage in region II, power law was followed with n = 1.2, 1.8, and 1.5 for d = ~ 4 nm, ~ 10 nm, and ~ 13 nm, respectively, which is associated with SCLC conduction mechanism. Further, at higher voltages, traps are activated in region III with slope n > 2. The n value for all thickness in positive and negative bias voltage are shown in **Table S3**.

Besides DC-based electrical measurements, frequency-dependent electrical studies (AC-based) were made for all thickness of MJs using impedance spectroscopy (EIS) in order to determine individual electrical parameters of the junctions, such as contact resistance, capacitance, and charge transfer resistance, which is equally important for predicting precise charge transport. The EIS measurements were done over a frequency range of $10^4$ Hz to 1 Hz by applying an AC bias 50 mV, while DC voltage was kept zero. To maintain a signal-to-noise (S/N) ratio, such high amplitude AC bias voltage was applied. **Figure 3e** shows a single depressed semicircle in the Nyquist plot indicating one capacitor. With an increase in thickness of Ru(Phen)$_3$ layers, the diameter of the semicircle also increases linearly, suggesting an increase in charge transfer resistance ($R_{ct}$) with the thickness. A similar observation was also observed in the Bode plot (**Figure 3f**). At low frequency, the |Z| is nearly constant (dominated by resistance of molecular layers) and decreases with the increase in frequency above the transition frequency ($f_T$).[51,52] In order to model our experimental impedance data, we used modified Randles equivalent circuit modelling (**Figure S21**) which consists of an uncompensated resistance, $R_u$ (mainly contact resistance in solid-state devices), in series with a parallel constant phase element (CPE) and a molecular layer resistance ($R_{ct}$), often modeled as charge-transfer resistance.[51] Details of EIS studies are shown in SI (**Figure S25, Table S4**).

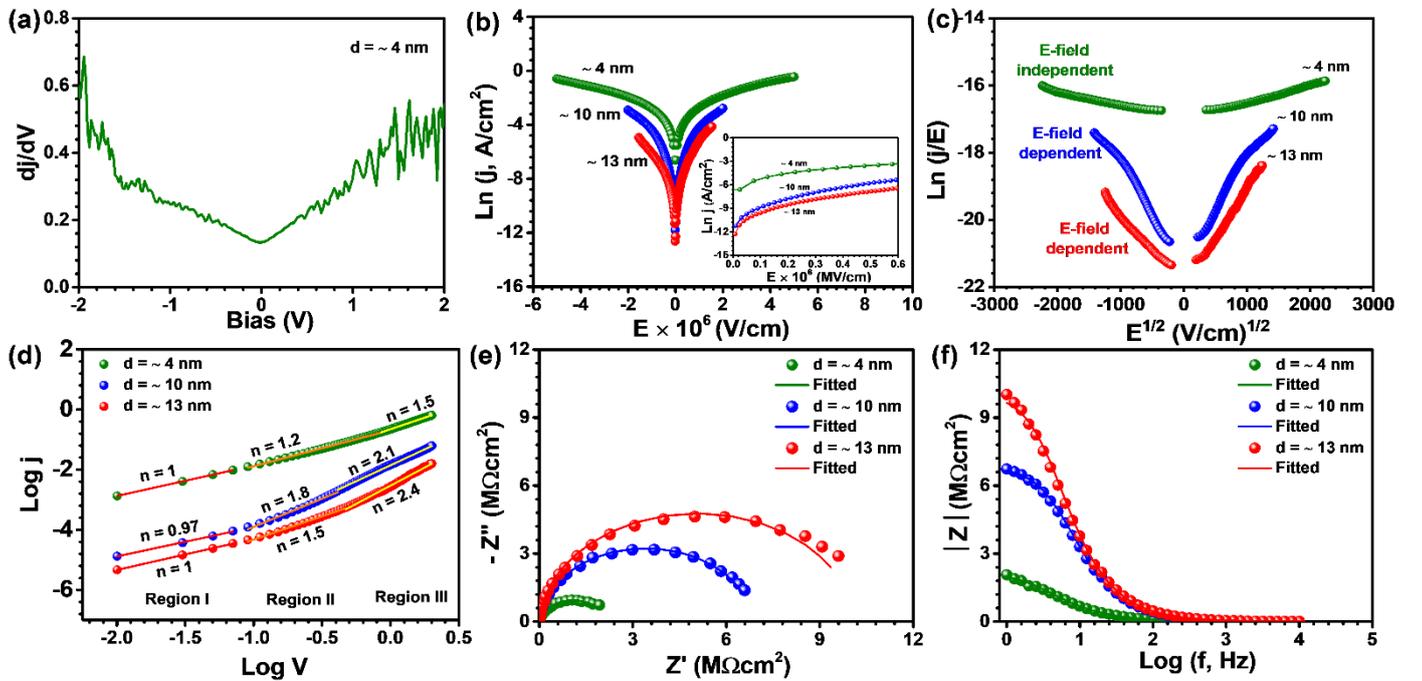

Figure 3. (a) dj/dV plot for Ru(Phen)$_3$ MJ for ~ 4 nm thickness. (b) Ln j vs. E plot for different thickness of Ru(Phen)$_3$ MJ. (c) PF plot and (d) Log j vs. Log V plot for MJs for different thickness of Ru(Phen)$_3$ MJ. (e) Nyquist and (f) Bode plot of Ru(Phen)$_3$ MJs.

**Calculations of isolated molecules**

We have performed computational study with (Ru(Phen)$_3$)$^{2+}$ complex and its trimer (Ru(Phen)$_3$)$_3^{6+}$. Geometry optimization and calculations of molecular orbitals were performed with B3LYP functional[53,54] and mixed electronic basics included 6-311G*[55] and lanl2dz[56] functions for ligand and metal atoms, respectively. UV spectra were defined with CAM-B3LYP functional[57] and the same basic set. Grimme's D3 corrections[58] were included to account possible non-covalent attraction between aromatic rings. Twenty excited states were considered. Time-dependent DFT (TD-DFT) approach and Tamm-Dancoff approximation implemented in TeraChem package were used.[59] Molecular structures and frontier orbitals of the (Ru(Phen)$_3$)$^{2+}$ and (Ru(Phen)$_3$)$_3^{6+}$ complexes are presented in **Figure 4a-c**. More orbitals of the (Ru(Phen)$_3$)$^{2+}$ are shown in supplementary **Figure S26**. One can see that the HOMO is located mostly on metal ion, whereas the LUMO is distributed on the ligands. Note that the difference between HOMO and HOMO – 1 orbitals as well as the difference between LUMO + 1 and LUMO orbitals is only about 0.1 eV. Calculated HOMO-LUMO gaps are equal to 3.64 and 2.66 eV for (Ru(Phen)$_3$)$^{2+}$ and (Ru(Phen)$_3$)$_3^{6+}$, respectively. Reducing the width of the HOMO-LUMO gap during the transition from a monomer to a polymer is a quite common effect. In the trimer, three orbitals of similar energy appear, each localized mainly on one of the ruthenium atoms. One of such orbitals is HOMO for the trimer. The UV absorbance spectra are shown in Figure S27 and excitation energies corresponding oscillators strengths are shown in Table S6. The main excitation energy of (Ru(Phen)$_3$)$_3^{6+}$ (364.9 nm) is in remarkable agreement with the experimental peak (365 nm).

Charge transport calculations

Transport properties of the (Ru(Phen)$_3$)$_3^{6+}$ complex were defined with the DFT combined with the non-equilibrium Green function approach implemented in the Quantum Espresso software.[60,61] This approach implements that the scattering region is placed between two semi-infinite leads. The left lead, ITO, was simulated with the In$_{28}$Sn$_4$O$_{48}$ cubic cell with the equilibrium lattice parameter $a_1$ = 1.030 nm.[62] The right aluminium lead was simulated with the Al$_4$ cubic cell with the equilibrium lattice parameter $a_2$ = 0.405 nm. Scattering region was represented by complex system, included 2x2x1 left lead cells and 5x5x1 right lead

cells connected by (Ru(Phen)$_3$)$_3$$^{6+}$ complex as shown in Figure 4d. Such molecular thin films orientation could be one of the possibilities. All calculations were performed with ultrasoft pseudopotentials[63] with cutoff energies of 60 and 360 Ry for wave functions and electron density, respectively. Taking into account the large system size, we use a 2x2 k-points grid[64] in the plane perpendicular to the transport direction. The transmission function is presented in Figure 4e. I-V curve presented in Figure 4f was calculated at temperature T = 300 K from transmission with the Landauer-Buttiker formula,[65]

$$I = \frac{2e}{h} \int_{-\infty}^{\infty} T(E)(F(E - 0.5eV) - F(E + 0.5eV))dE$$

Here $e$ and $h$ are the elementary charge and the Plank constant, respectively; $F(x) = 1/(1+\exp(x/kT))$ is the Fermi-Dirac distribution function. Fig. 4e shows that the transmission at Fermi level is low because it lies inside the gap of the molecule. Two peaks on $T(E)$ plot shows that at energy which is about 0.8 eV lower/higher the Fermi level the transmission is more probable due to resonance tunnelling through HOMO/LUMO orbitals. These peaks are the origin of the non-linear I-V characteristic presented at Fig. 4f. It demonstrates significant increase in the slope when the absolute value of V achieves about 0.8 eV.

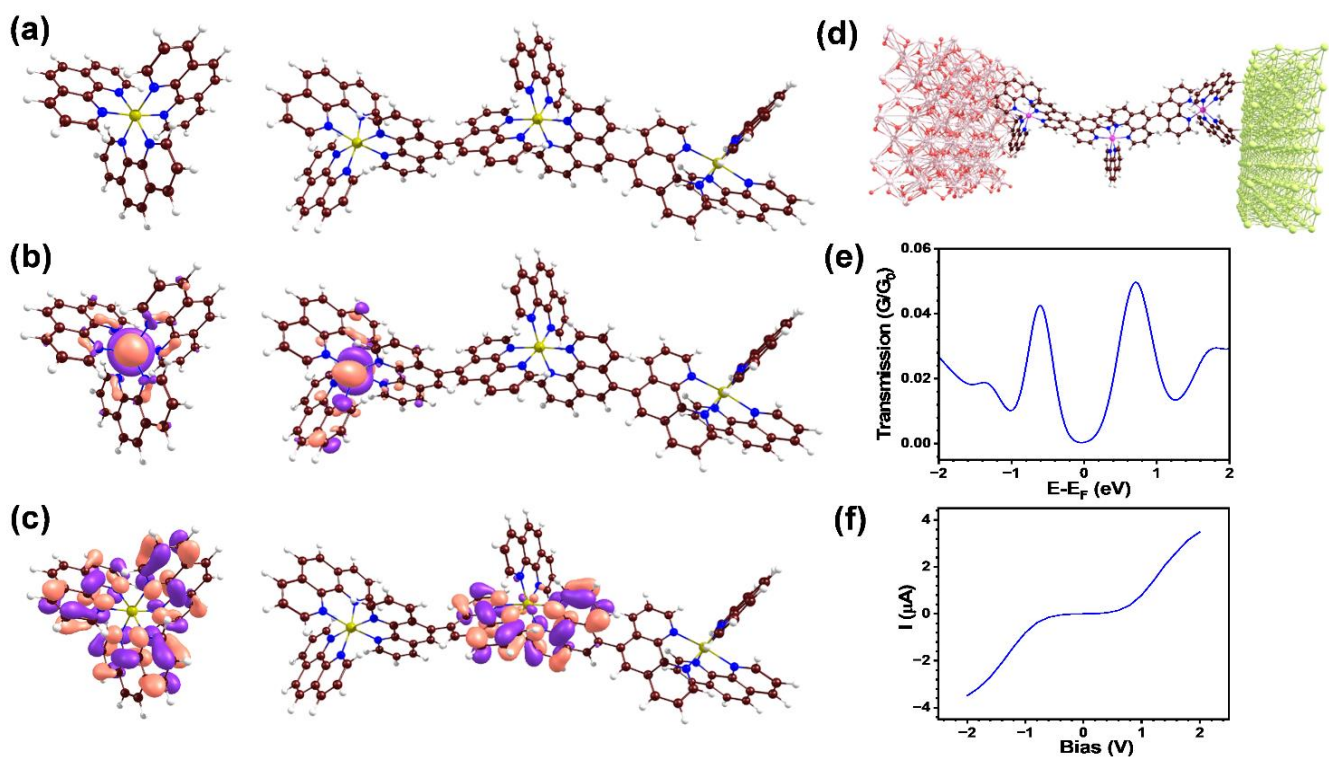

Figure 4. Molecular structure (a), HOMO (b) and LUMO (c) orbitals of the (Ru(Phen)$_3$)$^{2+}$ (left) and (Ru[Phen]$_3$)$_3$$^{6+}$ (right) complexes. (d) Atomistic model of the considered molecular junction. The (Ru[ligand]$_3$)$_3$$^{6+}$ complex connects ITO lead (left) with the aluminium lead (right). (e) Transmission, (f) I-V characteristic of the considered molecular junction.

Plausible charge transport mechanism

Based on the experimental and theoretical framework, a plausible mechanism for understanding the charge conduction of the devices containing Ru(Phen)$_3$ oligomers is illustrated in Figure 5. Considering a thickness of the single molecular unit which is approximately 1.3 nm (**Figure S28**), seven repeating units are placed (for 10 nm thick oligomer films) between two contacts. From the optical and electrochemical thin film data, HOMO and LUMO were estimated to be around − 5.62 eV and − 3.98 eV with a band gap of 1.64 eV, while $E_F$ of ITO and Al is − 4.7 eV and − 4.3 eV, respectively. Also, due to the covalent bonding of Ru(Phen)$_3$ with ITO electrode, we infer strong electronic coupling at ITO/Ru(Phen)$_3$ interface, and hence LUMO of Ru(Phen)$_3$ layer is expected to be pinned to the electrode and considerably broadened as shown in Figure 5a at no bias condition. When a bias voltage is applied (irrespective of bias polarity), electron injection into the layer will always be straightforward. Since the LUMO energy for the Ru(Phen)$_3$ layer is near the electrodes' Fermi ($E_F$) level, thus the electron-driven charge transport is involved (Figure 5b). Now the dominant charge transport will depend on thickness of Ru(Phen)$_3$ layers, and interachain hopping can occur above d > 4 nm. However, there are negatively charged $PF_6^-$ ions are present in the film, that can interact with the positively biased electrode surface, thus the potential profile might get shifted from the ideal linearity (Figure 5c).

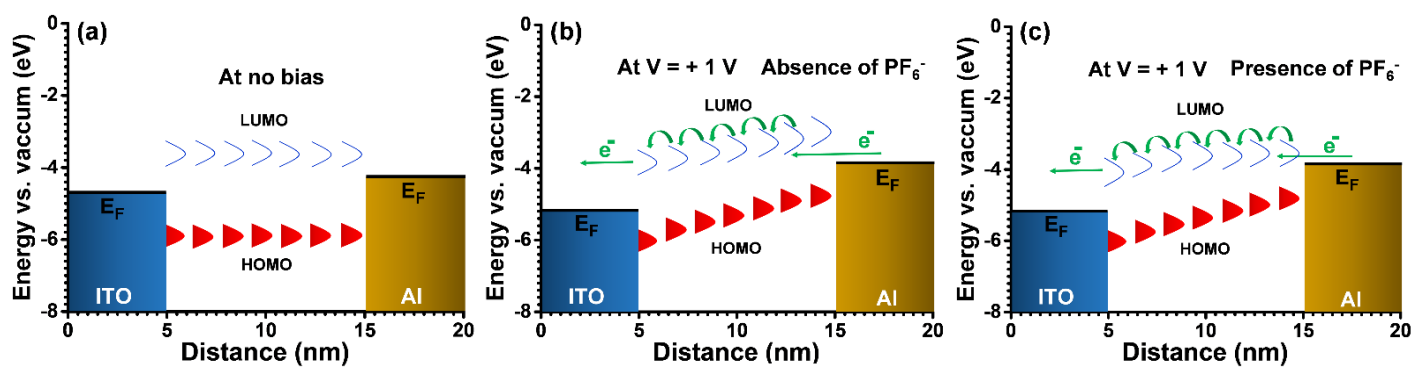

Figure 5. Energy profile for ITO/Ru(Phen)$_3$/Al MJs in the connected state (a) at no bias, (b) at + 1 V bias, (c) at + 1 V in presence of $PF_6^-$ counter ions.

**Conclusion**

The present work illustrates grafting of redox-active tris-diazonium salts of Ru(II)-phenanthroline via electrochemical reduction forming covalent bonded at patterned ITO/molecules interfaces, controlled thickness and three-dimensional utilized for molecular electronics. Such technique is crucial for fabricating high yield devices (**Table S6**). The work bears directly on pinpointing the charge transport mechanism for 3D Ru(Phen)$_3$ molecular layers with thickness between a limit of 4 to 13 nm. The large area ITO/Ru(Phen)$_3$/Al MJs showed efficient charge transport with a low β value of 0.48 nm$^{-1}$, no rectification at any thickness, and also E-field-dependent charge conduction (for thicker films), a first of a kind with such 3D redox active framework. We earmark these results to strong electronic coupling at ITO/Ru(Phen)$_3$ interface, Ru(Phen)$_3$ LUMO level dwelling between Fermi energy of the electrodes and degree of $PF_6^-$ counter ions present in the film, which makes charge transport facile inside the solid-state MJs. Overall, the work provides one basis for the rational design of 3D molecular structures for molecular devices with charge transport hallmarks.

**Acknowledgements**

RG thanks IIT Kanpur for a senior research fellowship for pursuing her Ph.D. program. PCM acknowledges financial support from Science and Engineering Research Board (SERB, Grant No. CRG/2022/005325), New Delhi, India. The authors acknowledge IIT Kanpur for infrastructures and equipment facilities.

**Conflict of Interest**

The authors declare no conflict of interest to this work.